# Coulomb-mediated single-electron heat transfer statistics across capacitively coupled silicon nanodots

Kensaku Chida[1,*], Antoine Andrieux[1], and Katsuhiko Nishiguchi[1]

[1]Basic Research Laboratories, NTT, Inc., 3-1 Morinosato Wakamiya, Atsugi, Kanagawa 243-0198, Japan.

*email: kensaku.chida@ntt.com

**Abstract**

Heat transfer mediated by the Coulomb interaction reveals unconventional thermodynamic behavior and broadens thermodynamics research into fields such as quantum dynamics and information engineering. Although some experimental demonstrations of phenomena utilizing Coulomb-mediated heat transfer have been reported, estimations of their performance, such as efficiency, and their theoretical evaluations necessitate qualitative evaluation of the transfer mechanism itself, which remains challenging. We present an experiment investigating single-electron dynamics in two electrostatically coupled silicon nanodots to quantify Coulomb-mediated heat transfer at the nanoscale. By estimating the Coulomb interaction strength between the dots using the cross-correlation measurements of the single-electron dynamics, we convert the single-electron dynamics into the statistics of Coulomb-mediated heat transfer. Conducting the experiment at equilibrium enabled us to obtain a fluctuating net-zero heat transfer between the dots. These heat transfer statistics are essential for exploring device functionalities from the perspective of stochastic thermodynamics and for verifying universal relations in nonequilibrium states.

**Introduction**

In heat-driven nanodevices, such as thermoelectric components, electrons simultaneously convey charge and heat, making it difficult to separate the charge and heat currents. This means that the performance of thermoelectric devices is limited by both thermal conductivity and electrical conductivity. In contrast, when nanodevices or quantum dots are brought into close proximity, it becomes possible to transfer heat without an accompanying charge flow, because the energy exchange is mediated by Coulomb interactions between electrons [1-3]. Utilizing this type of heat transfer in nanodevices has enabled experimental demonstrations of multi-terminal heat engines [4-6] and information heat engines [7]. These experiments have revealed unconventional heat engine operations that fall outside the theoretical framework used to evaluate the performance of conventional two-terminal thermoelectric devices [8, 9], where electric current and heat current flow simultaneously in the same location. This has spurred development and growth of research in quantum thermodynamics [10, 11] and information thermodynamics [12-21], which aim to expand the scope of the thermodynamic framework to include nonequilibrium states.

Although unconventional heat engine operations driven by Coulomb-mediated heat transfer in nanodevices have been demonstrated, direct measurement of this heat transfer remains unachieved. If this heat transfer could be observed at the trajectory level, the statistical information that could be gathered on it would enable analyses from the viewpoint of stochastic thermodynamics [22-28]. This would allow us to gain precise and unified knowledge on the efficiency, accuracy, and speed of nanodevice operations [29, 30]. Moreover, the statistics would provide essential information to verify universal relations in nonequilibrium states [31-38]. However, measurement of heat transfer statistics at the nanoscale remains technically challenging.

A method based on single-electron detection has been theoretically proposed to measure the Coulomb-mediated heat transfer statistics [3]. If the Coulomb interaction strength $E_{Cm}$ between the dots is known, observing the series of electron transitions indicated by 1 to 4 in Fig. 1(a) is equivalent to observing the heat transferred $Q = E_{Cm}$ from Dot 1 to Dot 2. The mechanism of this heat transfer can be explained by the contribution of the Coulomb interactions to the chemical potential of the dots. The chemical potential of Dot 1(2) is expressed as

$$\mu_{1(2)}(N_1, N_2) = 2N_{1(2)}E_{C1(2)} + N_{2(1)}E_{Cm} + f_{1(2)},$$

where $N_{1(2)}$ is the relative number of electrons in Dot 1(2), $E_{C1(2)}[= e^2/2C_{1(2)}]$ is the charging energy of Dot 1(2), $e$ is the elementary charge, $C_{1(2)}$ is the total capacitance of Dot 1(2), and $f_{1(2)}$ is a constant independent of $N_1$ and $N_2$. The first term represents the change in $\mu_{1(2)}$ due to the confinement of single electrons within the dot, whereas the second term represents the change in $\mu_{1(2)}$ caused by the Coulomb interaction between the dots. Due to the second term, when $N_{2(1)}$ varies as an electron enters or exits Dot 1(2), $\mu_{1(2)}(N_1, N_2)$ changes by an integer multiple of $E_{Cm}$ between the entry and exit of the electron. This results in the Coulomb-mediated heat transfer between the two dots. In the sequence of electron transitions labeled 1 to 4 in Fig. 1(a), an electron entering Dot 1 loses energy equivalent to $E_{Cm}$ before returning to the electron reservoir (ER), while an electron entering Dot 2 gains energy equivalent to $E_{Cm}$ before returning to the ER. The $E_{Cm}$ lost at Dot 1 is transferred to Dot 2, resulting in an overall heat transfer $Q = E_{Cm}$ from Dot 1 to Dot 2. Therefore, evaluation of the single-electron dynamics in the two dots is crucial for observing this heat transfer, and $E_{Cm}$ serves as a conversion factor that translates the single-electron dynamics into heat transfer statistics [1-3].

In this paper, we present a measurement of single-electron thermal fluctuations [39] in capacitively coupled silicon nanodots to ascertain the heat transfer statistics between the two dots. To achieve this, we estimate $E_{Cm}$ by using the cross-correlation measurements [40-46] of the thermal fluctuations in the two dots. Despite the stochastic nature of the thermal fluctuations in each dot, the repulsive Coulomb interaction between the dots induces a negative correlation in the thermal fluctuations of the individual electrons entering and exiting the two dots. We derive $E_{Cm}$ from the cross-correlation function between the single-electron thermal fluctuations in the two dots. Then, by using $E_{Cm}$ as a conversion factor, we convert the single-electron dynamics into the Coulomb-mediated heat transfer statistics. At equilibrium, we obtain statistics on the fluctuating net-zero heat transfer between the two dots. Measuring heat transfer statistics will benefit the development of nonequilibrium physics, including quantum thermodynamics and information thermodynamics. By studying these statistics under nonequilibrium conditions, we can conduct precise analyses of device functions induced by this heat transfer, such as for information heat engines, within the framework of stochastic thermodynamics.

**Results**

**Device structure and observation of single-electron thermal motion.** Figure 1(b) shows an electron microscope image of the device. The device was fabricated on a silicon-on-insulator (SOI) wafer and had two nanometer-scale dots (Dot 1 and Dot 2) connected to a common electron reservoir (ER). The two dots were separated by approximately 20 nm of $SiO_2$, and while they were electrostatically coupled, they were electrically insulated. A three-terminal charge detector, located 20 nm away from the dots, was used to detect the change in the number of electrons [$N_1(t)$ and $N_2(t)$] in the two dots. The upper gate (UG) was used to induce electrons in the device, and the lower gate (LG) was used to form an energy barrier between the dots and the ER. By adjusting the negative voltage applied to the LG, the transition rates of single electrons thermally hopping between the ER and dots can be modulated. All experiments were performed at room temperature (300 K).

By analyzing the current flowing through the three terminals D1, D2, and D3 [$I_{D1}(t)$, $I_{D2}(t)$, and $I_{D3}(t)$, respectively] of the three-terminal charge detector [Fig. 1(c)-(e)], we determined the number of electrons in each of the two dots [Fig. 2(a), (b)] and then performed electron counting statistics on the results [46-49]. Note that the values of $N_1(t)$ and $N_2(t)$ are relative, defined as zero at the start of the measurement, and are not the absolute numbers of electrons in the dots. The random changes in $N_1(t)$ and $N_2(t)$ are caused by thermal hopping of single electrons between the ER and dots, and the probability distributions $p(N_1)$ and $p(N_2)$ appear to be independent Gaussian distributions. This indicates that $E_{Cm}$ is sufficiently small compared to $E_{C1}$ and $E_{C2}$.

As the thermal fluctuations in each dot are completely random, no information about the interactions between the two dots can be obtained from the electron dynamics in one dot. However, their joint probability distribution $p(N_1, N_2)$ contains information about the interactions between the dots. If there are interactions between the two dots, $p(N_1, N_2)$ becomes an elliptically asymmetric distribution tilted with respect to the $N_1$ and $N_2$ axes, with reduced probabilities in the first and third quadrants, indicating negative correlation between $N_1$ and $N_2$: the repulsive Coulomb interaction suppresses simultaneous occupation and vacancy of the two dots [Fig. 2(c)]. Although the tilted angle of this elliptical distribution provides information on the correlation affecting $E_{Cm}$, it is difficult to estimate it because the joint probability with the large discreteness of the $N_1$ and $N_2$ axes makes this theoretical fitting difficult. For this reason, we will instead discuss the correlation function of $N_1(t)$ and $N_2(t)$.

**Correlation functions of single-electron thermal fluctuations.** The auto-correlation functions $C_{\alpha\alpha}(\tau)$ and cross-correlation functions $C_{\alpha\beta}(\tau)$ of $N_1(t)$ and $N_2(t)$ are defined as

$$C_{\alpha\beta}(\tau) = \frac{1}{n_{\text{meas}}} \sum_{n=0}^{n_{\text{meas}}-1} N_\alpha^{(N)}(t) N_\beta^{(N)}(t+\tau),$$

where $\alpha, \beta = 1$ or 2, $\tau$ is the time delay, $n$ is the measurement number in the time-series measurement, and $n_{\text{meas}}$ is the total number of measurements. $N_\alpha^{(N)}(t) = \frac{N_\alpha(t) - \mu_\alpha}{\sigma_\alpha}$ is the normalized $N_\alpha(t)$, where $\mu_\alpha$ is the mean and $\sigma_\alpha$ is the standard deviation of $N_\alpha(t)$. When $\alpha = \beta$, the auto-correlation function is obtained, and when $\alpha \neq \beta$, the cross-correlation function is obtained.

The auto-correlation functions $C_{11}(\tau)$ and $C_{22}(\tau)$ decay nonlinearly to zero, reflecting that single-electron motion is a random phenomenon caused by thermal fluctuation [Fig. 3(a)]. By fitting with a single exponential function, we found the relaxation times $\tau_{C11}$ and $\tau_{C22}$ of $C_{11}(\tau)$ and $C_{22}(\tau)$ to be 5.4 s and 9.8 s, respectively. As we expected the relaxation to originate from thermal hopping between the ER and dots, we decided to examine the consistency between the relaxation times and the electron transition rates of the thermal hopping.

The electron transition rates from the ER to the dots, $\Gamma_{N_{1(2)}}^{\text{in}}$, and from the dots to the ER, $\Gamma_{N_{1(2)}}^{\text{out}}$, are

defined as follows: [39]

$$\Gamma_{N_{1(2)}}^{\text{in/out}} = \frac{1}{\langle \tau_{N_{1(2)}} \rangle} \left( \frac{P_{N_{1(2)}}^{\text{in/out}}}{P_{N_{1(2)}}^{\text{in}} + P_{N_{1(2)}}^{\text{out}}} \right),$$

where $\langle \tau_{N_{1(2)}} \rangle$ is the average dwell time of Dot 1(2) in the state $N_{1(2)}$, $P_{N_{1(2)}}^{\text{in}}$ is the transition probability from $N_{1(2)}$ to $N_{1(2)}+1$, and $P_{N_{1(2)}}^{\text{out}}$ is the transition probability from $N_{1(2)}$ to $N_{1(2)}-1$. These parameters were obtained from the time-series data of $N_{1(2)}$ [Fig. 1(c)-(e)]. The electron transition rates $\Gamma_{N_1(N_2)}^{\text{in}} = \Gamma_{\text{P1(P2)}}$ from the ER to the dots are independent of $N_{1(2)}$, while the electron transition rates $\Gamma_{N_1(N_2)}^{\text{out}} = \Gamma_{\text{P1(P2)}} \exp\{-[E_{\text{C1(C2)}} N_{1(2)}]/kT\}$ from the dots to the ER show an exponential dependence on $N_{1(2)}$ [Fig. 3(b) and 3(c)] [39]. Here, $\Gamma_{\text{P1(P2)}}$ is the prefactor of thermal hopping, $kT = 26$ meV represents the thermal energy at $T = 300$ K, and $k$ is the Boltzmann constant. $\Gamma_{\text{P1(P2)}}$ corresponds to the transition rate at the point where $\Gamma_{N_1(N_2)}^{\text{out}}$ and $\Gamma_{N_1(N_2)}^{\text{in}}$ intersect in Fig. 3(b) and 3(c), representing the frequency of thermal hopping. We obtained a $\Gamma_{\text{P1}}$ of 0.39 s$^{-1}$ and $\Gamma_{\text{P2}}$ of 0.15 s$^{-1}$, indicating that the electron transition at Dot 1 is about twice as fast as at Dot 2. This is consistent with the approximately two-fold faster relaxation at Dot 1 compared to Dot 2. Furthermore, Monte Carlo simulations of the thermal hopping between the ER and dots, using only the parameters solely derived from the experiment, replicated the experimentally observed $C_{11}(t)$ and $C_{22}(t)$ [blue lines in Fig. 3(a)]. These findings from the experimental data and simulations confirm that the observed single-electron dynamics originated from thermal hopping.

We can investigate the interaction between the two dots by examining the cross-correlation functions. The cross-correlation function $C_{\alpha\beta}(\tau)$ of $N_1(t)$ and $N_2(t)$ reflects the repulsive Coulomb interaction between electrons in the two dots, showing negative values for a time difference $\tau < 20$ s [Fig. 4(a)]. $C_{12}(\tau)$ [$C_{21}(\tau)$] exhibits a minimum value of approximately -0.12 at $\tau = 0$ s and its magnitude monotonically decays to zero with a relaxation time $\tau_{C_{12}(C_{21})}$ of ~8 s (~14 s). These relaxation times were comparable to those of the autocorrelation functions. This in turn suggests that the correlation between the dots induced by the Coulomb interaction disappears as a result of thermal fluctuations causing electron transitions.

It is noteworthy that the absolute value of $C_{\alpha\beta}(0)$ reflects $E_{\text{Cm}}$ between the two dots. To estimate $E_{\text{Cm}}$, we performed Monte Carlo simulations with parameters obtained from the experiments, except for $E_{\text{Cm}}$. From the simulations with various $E_{\text{Cm}}$, we found that $C_{\alpha\beta}(0)$ monotonically decreases as $E_{\text{Cm}}$ increases [Fig. 4(b) and 4(c)]. By comparing the simulations and experimental results, we estimated the strength of the Coulomb interaction between dots to be $E_{\text{Cm}} \sim 2$ meV [Fig. 4(c)]. That is, the simulation with $E_{\text{Cm}} = 2$ meV reproduced $C_{\alpha\beta}(\tau)$ found in the experiments [the blue lines in Fig. 4(a)]. This corroborates the finding that the negative correlation is caused by the repulsive Coulomb interaction between the dots.

**Observation of the Coulomb-mediated single-electron heat transfer statistics between the dots.** Having quantified $E_{\text{Cm}}$ by using cross-correlation measurements, we turned to estimating the Coulomb-mediated heat transfer from single-electron dynamics on the basis of the theoretical proposal [3]. Here, we obtained the statistics of heat transfer between the two dots from the trajectory of [$N_1(t)$, $N_2(t)$]. The rotational direction of a closed loop drawn by the trajectory of [$N_1(t)$, $N_2(t)$] on the $N_1$–$N_2$ state space correspond to the direction of the Coulomb-mediated heat transfer: a counterclockwise (clockwise) closed loop results in $Q$ flowing from Dot 1 (Dot 2) to Dot 2 (Dot 1). On the $N_1$–$N_2$ state space, the series of electron transitions (0, 0) → (0, 1) → (1, 1) → (1, 0) → (0, 0) that transfer $Q = E_{\text{Cm}}$ forms a counterclockwise closed loop with an area of 1 and its time-reversed series of electron transitions (0, 0) → (1, 0) → (1, 1) → (0, 1) → (0, 0) that transfers $Q = -E_{Cm}$ forms a clockwise closed loop [Fig. 5(a)]. The sign of $Q$ defined as positive when the flow is from Dot 1 to Dot 2.

The theoretical proposal [3] considered the Coulomb blockade regime, where $N_1$ and $N_2$ are restricted to 0 and 1. However, we will show that by analyzing the trajectory of $[N_1(t), N_2(t)]$ on the $N_1$-$N_2$ state space, the proposed method is capable of determining the Coulomb-mediated heat transfer even when the Coulomb blockade is lifted and multiple electrons are involved. When multiple electrons take part in forming a closed loop, the amount of $Q$ associated with a loop corresponds to the product of the area $S$ enclosed by the loop and $E_{Cm}$: $Q = SE_{Cm}$. For example, the series of electron transitions (0, 0) → (0, 2) → (1, 2) → (1, 0) → (0, 0) forms a counterclockwise loop with $S = 2$ and transfers $Q = 2E_{Cm}$ [Fig. 5(b)]. In the most frequent case, the area enclosed by a loop is zero ($S = 0$) and the loop is composed only of pairs of electron transitions that cancel each other out in terms of heat transfer, resulting in $Q = 0$ [Fig. 5(c)].

The decomposition of the trajectory $[N_1(t), N_2(t)]$ on the $N_1$–$N_2$ state space into closed loops enables us to obtain the statistics of the Coulomb-mediated heat. The trajectory shown in Fig. 5(d) can be decomposed into a total of five closed loops: one counterclockwise loop with an area of 1, one clockwise loop with an area of 3, and three zero-area loops, corresponding to Q = $E_{Cm}$, -3$E_{Cm}$, and 0, respectively. By performing the decomposition on the experimentally obtained trajectory $[N_1(t), N_2(t)]$, we obtained a net zero heat transfer corresponding to the equilibrium state of the two dots [Fig. 5(e)]. The probability distribution $P(Q)$ derived from the experimentally obtained trajectory is symmetric with respect to $Q = 0$, indicating that even in equilibrium, there are finite fluctuations in the Coulomb-mediated heat transfer due to thermal fluctuations. The distribution rapidly decays as the absolute value of $Q$ increases and appears to be an exponential distribution [50]. However, to discuss the probability distribution function, it will be necessary to evaluate the tails making extended and more precise measurements. Nonetheless, our Monte Carlo simulations conducted under the condition of $kT$ = 26 meV reproduced the experimentally obtained distribution of $P(Q)$ from the trajectory on the $N_1$–$N_2$ state space obtained [the blue line in Fig. 5(e)].

**Discussion**

As the trajectory on the $N_1$–$N_2$ state space originates from thermal hopping, the distribution of $P(Q)$ is expected to be temperature dependent. We investigated this idea by using Monte Carlo simulations and found that an increase (decrease) in $kT$ results in an increase (decrease) in $P(Q)$ for larger $Q$ and a broader (narrower) distribution, reflecting that the increase (decrease) in thermal energy leads to a larger (fewer) number of single electrons participating in forming a closed loop. This suggests that the heat transfer statistics are governed by thermal fluctuations. Additionally, the relaxation time of $C_{\alpha\beta}(\tau)$ and the transition rates $\Gamma_{P1}$, and $\Gamma_{P2}$ were independent of $E_{Cm}$, at least within the range of $E_{Cm}$ from 0 to 10 meV [see Fig. 4(b)]. This supports the idea that electron transitions caused by thermal hopping are the mechanism behind the disappearance of the correlation between the two dots. These results obtained with Monte Carlo simulations indicate that we can estimate the Coulomb-mediated heat transfer statistics as long as single-electron dynamics are detectable across a wide range of experimental conditions.

In summary, using single-electron detection, we obtained the statistics of the Coulomb-mediated heat transfer between electrostatically coupled silicon nanodots. The cross-correlation measurements of single-electron thermal fluctuations in the two dots enabled us to quantify the strength of the Coulomb interaction between them. By using this interaction strength as a conversion factor, we converted the single electron dynamics in the nanodots into the statistics of the heat transfer between them. At equilibrium, we observed a fluctuating net zero Coulomb-mediated heat transfer between them.

By driving our silicon nanodevices into nonequilibrium states [51-54], we can investigate device functions arising from the Coulomb-mediated heat current, such as thermal diodes [1-3]. Given that the method used in this study is applicable to multi-terminal heat engines and information heat engines, measuring the heat transfer statistics in nonequilibrium conditions allows for a precise and consistent discussion on nanodevice operation from the viewpoint of stochastic thermodynamics [22-30]. Moreover, the cross-correlation measurements enable us to delve into the foundations of nonequilibrium physics, encompassing universal relationships concerning the asymmetry of the cross-correlation function [31-35] and the fluctuation-response relationships in nonequilibrium steady states [36-38]. Direct measurement of Coulomb-mediated heat transfer statistics offers essential insights for studying the functionalities of nanodevices from the perspective of nonequilibrium physics, including stochastic thermodynamics and quantum thermodynamics.

**Acknowledgements**

We appreciate the fruitful discussions with Keiji Saito, Yasuhiro Tokura, Takaaki Monnai, Naruo Ohga, Motoki Asano, Toshiaki Hayashi, Takase Shimizu, Akira Fujiwara, and Hiroshi Yamaguchi. K.C. appreciates the help given by Gento Yamahata in fabricating the device.

**Methods**

**Device fabrication.** The device was fabricated from an SOI wafer. An ER, two dots, and a three-terminal detector were formed on an SOI layer with a boron concentration of $10^{15}$ cm$^{-3}$. The width and thickness of SOI channels for the dots and detector were approximately 30 nm and 20 nm, respectively. The oxide thickness was 9 nm. Then, an LG composed of polycrystalline silicon was formed on the SOI channel between the dots and ER. The width of the LG was 60 nm. The entire part was then covered with a 50-nm-thick oxide interlayer formed by chemical vapor deposition. Finally, a UG that covered the main part of the device was formed over the interlayer. The UG was used to induce electrons in the device and to control the current flowing through the detector channels. Details of the process are provided in ref. [55].

**Experiments.** All experiments were conducted at 300 K in a vacuum using a probe station with temperature control. A semiconductor device parameter analyzer (Keysight B1500A) was used to apply voltage to the electrodes and measure current through the three-terminal detector. To capture electron transitions between the ER and dots, the transition rate of the single electrons was adjusted to $\Gamma_{P1(P2)}$ ~1 Hz, which is about an order of magnitude slower than the measurement frequency $f_{\text{meas}} \approx 11$ Hz [56, 57].

In the analytical process to determine $N_1(t)$ and $N_2(t)$, we excluded signals from unintended sources other than single-electron movements into and out of the two dots through the following steps. First, we defined the change in $I_{Di}(t)$, where $i$ = 1, 2, or 3, due to a change in the number of electrons in Dot 1(2) as $\delta I_{Di,\text{Dot1(2)}}$. Owing to the different coupling strengths between the three detector terminals and the two dots for each combination of a terminal and a dot, the ratio $\delta I_{Di,\text{Dot1}}/\delta I_{Di,\text{Dot2}}$ varies for each terminal [Fig. 1 (c)-(e)]. By combining these signals, we were able to distinguish signals from the two dots from noise originating from external sources as follows. We excluded changes in $I_{Di}$ that could not be expressed as

$\Delta N_1(t)\delta I_{\mathrm{D}i,\mathrm{Dot1}} + \Delta N_2(t)\delta I_{\mathrm{D}i,\mathrm{Dot2}}$ and identified them as signals from sources other than the dots. Then, when the amplitude of signals in two or more $I_{\mathrm{D}i}(t)$ matched $\Delta N_1(t)\delta I_{\mathrm{D}i,\mathrm{Dot1}} + \Delta N_2(t)\delta I_{\mathrm{D}i,\mathrm{Dot2}}$ with the same $\Delta N_1(t)$ and $\Delta N_2(t)$, we adjusted the number of electrons in $N_1(t)$ by $\Delta N_1(t)$ and in $N_2(t)$ by $\Delta N_2(t)$. In cases where $\Delta N_1(t)\delta I_{\mathrm{D}i,\mathrm{Dot1}} + \Delta N_2(t)\delta I_{\mathrm{D}i,\mathrm{Dot2}}$ matched the experimental results within ±0.2 nA for two or more combinations of $\Delta N_1(t)$ and $\Delta N_2(t)$, we selected the $\Delta N_1(t)$ and $\Delta N_2(t)$ that minimized the distance in the $N_1$-$N_2$ state space. If $\Delta N_1(t)$ and $\Delta N_2(t)$ were both non-zero simultaneously, we considered the electron transitions to have occurred simultaneously. This procedure enabled us to distinguish signals from the two dots from those originating externally, such as electron movements into and out of charge traps near the detector.

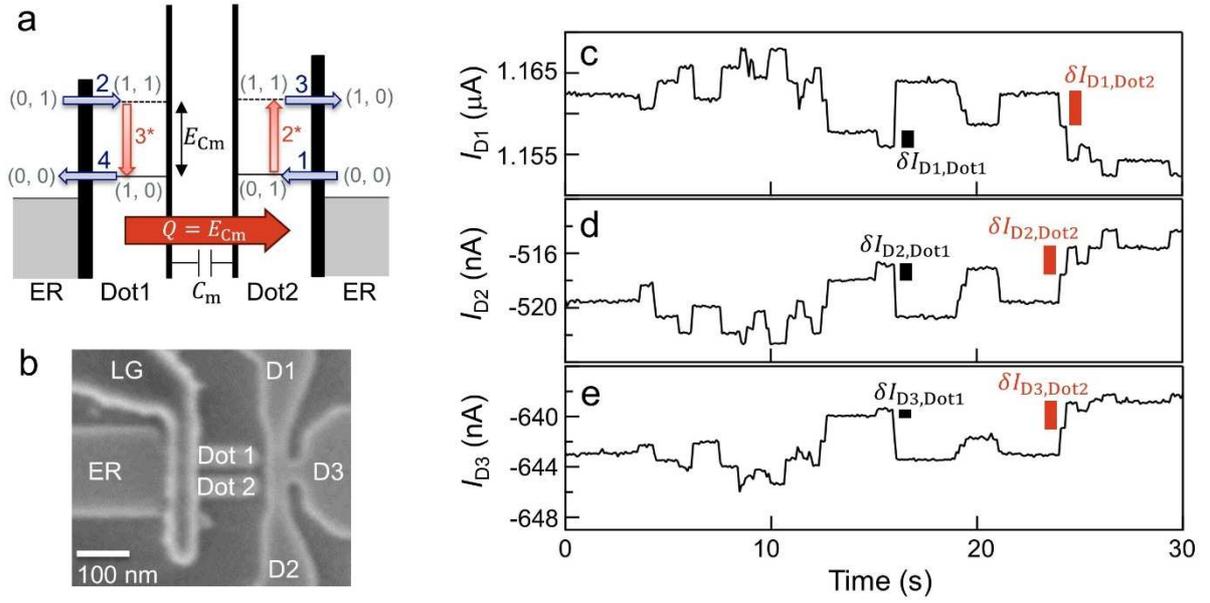

**Fig. 1. Device structure for observing Coulomb-mediated single-electron heat transfer.** (**a**) Schematic diagram of Coulomb-mediated single-electron heat transfer $Q$ across capacitively coupled nanodots. Sequential electron transitions (1 to 4) alter the number of electrons in the dots $(N_1, N_2)$ from $(0, 0) \rightarrow (0, 1) \rightarrow (1, 1) \rightarrow (1, 0) \rightarrow (0, 0)$, where $N_{1(2)}$ is the number of electrons in Dot 1(2). In transition 1 $[(0, 0) \rightarrow (0, 1)]$, an electron enters Dot 2. In transition 2 $[(0, 1) \rightarrow (1, 1)]$, an electron enters Dot 1, raising the chemical potential of Dot 2 $\mu_2(N_1, N_2)$ by $E_{Cm}$ (red upward arrow 2*). In transition 3 $[(1, 1) \rightarrow (1, 0)]$, an electron exits Dot 2, lowering the chemical potential of Dot 1 $\mu_1(N_1, N_2)$ by $E_{Cm}$ (red downward arrow 3*). Finally, in transition 4 $[(1, 0) \rightarrow (0, 0)]$, an electron exits Dot 1, returning the dots to their initial state. The key point is that an electron exits Dot 1(2) in transition 4(3) from $E_{Cm}$ lower (higher) $\mu_{1(2)}(N_1, N_2)$ compared with $\mu_{1(2)}(N_1, N_2)$ when an electron enters Dot 1(2) in transition 2(1). Overall, the $E_{Cm}$ lost in Dot 1 is gained in Dot 2, resulting in heat transfer $Q = E_{Cm}$ from Dot 1 to Dot 2, despite no charge transport between the dots. (**b**) Scanning electron microscope image of the device structure. Two 100-nm-long dots (Dot1 and Dot2) and a detector with three terminals (D1, D2, and D3) are electrically insulated by a 20-nm-wide trench filled with $SiO_2$. The dots are connected to the electron reservoir (ER) via a 60-nm-long polycrystalline-silicon lower gate (LG) electrode. The width and height of the dots and the detector channels are about 30 nm and 20 nm, respectively. The upper gate (UG) is formed on a 50-nm-thick interlayer of $SiO_2$. (**c-e**) Current $I_{Di}$ through the detector terminals D$i$ as a function of time, where $i$ =1, 2, or 3. The black (red) bars indicate the amplitude of the single-electron signal from Dot 1(2) $\delta I_{Di,Dot1}$ ($\delta I_{Di, Dot2}$). The ratios $\delta I_{D1,Dot1}/\delta I_{D1, Dot2}$, $\delta I_{D2,Dot1}/\delta I_{D2, Dot2}$ and $\delta I_{D3,Dot1}/\delta I_{D3, Dot2}$ are 1/2, 3/5, and 3/10, respectively.

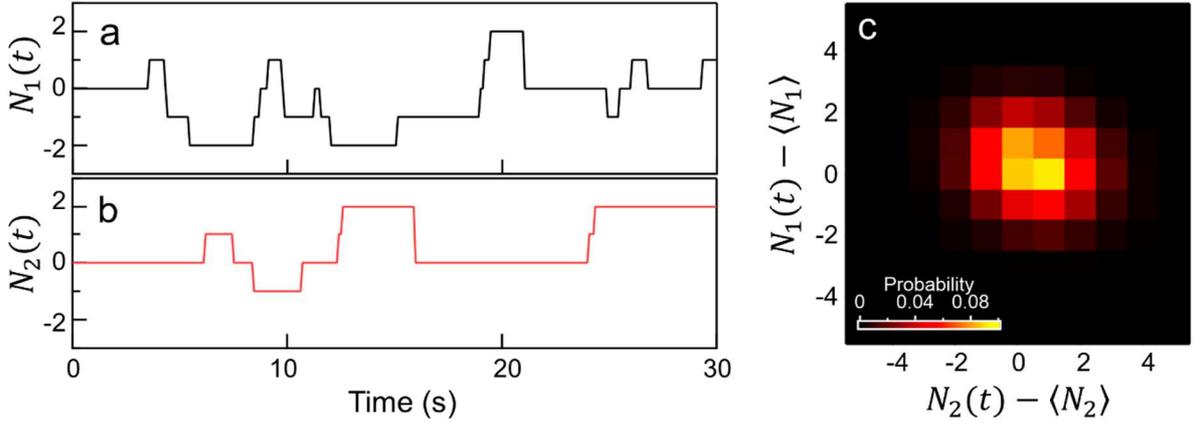

**Fig. 2. Single-electron motion in the capacitively coupled silicon dots.** (**a**) $N_1(t)$ and (**b**) $N_2(t)$ as a function of time obtained from a combination of $I_{D1}(t)$, $I_{D2}(t)$, and $I_{D3}(t)$. (**c**) The joint probability distribution $P[N_1(t), N_2(t)]$ obtained from the data in Fig. 2(a) and (**b**), where $\langle N_1 \rangle$ and $\langle N_2 \rangle$ are the average of $N_1(t)$ and $N_2(t)$, respectively. Due to an interaction between the dots, $p(N_1, N_2)$ is a tilted asymmetric elliptical distribution with respect to the $N_1$ and $N_2$ axes. From the probability distribution, we estimated $E_{C1(C2)}$ as 7.0 meV (9.0 meV) [39]. Note that to obtain $E_{C1(C2)}$, it is necessary to use the conditional probability distribution $p(N_{1(2)}|N_{2(1)}=N)$ when $N_{2(1)}$ is at a specific value $N$. Otherwise, as $p[N_{1(2)}]$ includes information on the thermal fluctuations of $N_{2(1)}$ in the other dot, the estimated variance will be larger than the actual value and $E_{C1(C2)}$ will be underestimated.

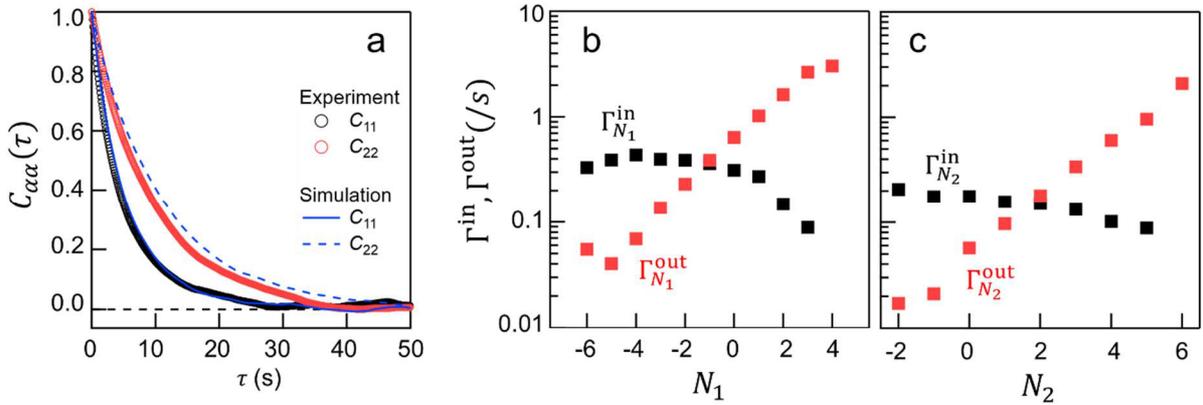

**Fig. 3. The auto-correlation function of thermal fluctuation at each dot.** (**a**) Auto-correlation function $C_{\alpha\alpha}(\tau)$ of $N_\alpha$ as a function of $\tau$, where $\alpha = 1$ or 2. The blue solid (dashed) line is $C_{11}(\tau)$ [$C_{22}(\tau)$] obtained from Monte Carlo simulations of thermal hopping between the ER and the capacitively coupled dots. Electron transition rates at (**b**) Dot 1 and (**c**) Dot 2 as a function of $N_1$ and $N_2$, respectively. We determined $\Gamma_{P1(P2)}$ to be 0.39 s$^{-1}$ (0.15 s$^{-1}$) from the intersection of $\Gamma_{N_\alpha}^{\text{out}}$ and $\Gamma_{N_\alpha}^{\text{in}}$. Monte Carlo simulations were performed with the following parameters: measurement interval $dt = 0.1$ s, total measurement time $t_{\text{meas}} = 30{,}000$ s, $\Gamma_{P1} = 0.39$ s$^{-1}$, $\Gamma_{P2} = 0.15$ s$^{-1}$, $E_{C1} = 7$ meV, $E_{C2} = 9$ meV, and $kT = 26$ meV.

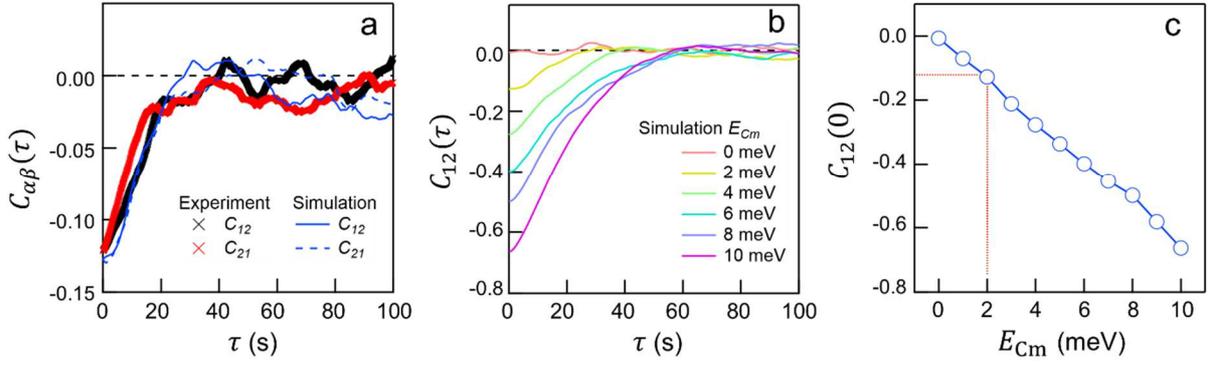

**Fig. 4. Cross-correlation function of single-electron dynamics at the two dots and the strength of the Coulomb interaction $E_{Cm}$.** (a) Cross-correlation function $C_{\alpha\beta}(\tau)$ of $N_\alpha(\tau)$ and $N_\beta(\tau)$ as a function of $\tau$, where $\alpha$, $\beta$ = 1 or 2. The blue solid (dashed) line is $C_{12}(\tau)$ [$C_{21}(\tau)$] obtained from a Monte Carlo simulation. Negative values of $C_{\alpha\beta}(\tau)$ with $\tau$ < 20 s indicate that the dynamics of the thermal motions of the single electrons in the nanodots are out of phase because of the repulsive Coulomb interaction. As the two dots are at equilibrium, $C_{12}(\tau) = C_{21}(\tau)$ within the fluctuation of $C_{\alpha\beta}(t)$ ~0.03 because of the limited number of measurement points. (b) Simulated $C_{12}(\tau)$ as a function of $\tau$ with $E_{Cm}$ from 0 to 10 meV. (c) Simulated $C_{12}(0)$ as a function of $E_{Cm}$. As shown in (a), $C_{12}(0)$ in this experiment is approximately -0.12, which corresponds to $E_{Cm}$ ~2 meV. The Monte Carlo simulations were performed with $dt$ = 0.1 s, $t_{meas}$ = 30,000 s, $\Gamma_{P1}$ = 0.39 s$^{-1}$, $\Gamma_{P2}$ = 0.15 s$^{-1}$, $E_{C1}$ = 7 meV, $E_{C2}$ = 9 meV, $kT$ = 26 meV, and $E_{Cm}$ = 2.0 meV if not otherwise indicated.

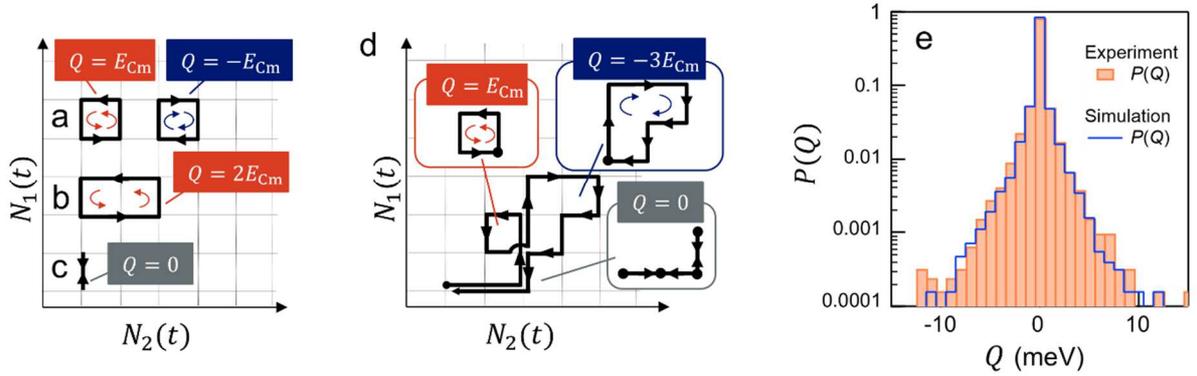

**Fig. 5. Coulomb-mediated single-electron heat transfer statistics.** (a-c) Schematic illustration of correspondence between the trajectory of [$N_1(t)$, $N_2(t)$] and the Coulomb-mediated heat transfer $Q$ across the dots. The direction of $Q$ is from Dot 1(2) to Dot 2(1) when the trajectory is counterclockwise (clockwise) on the state space, and the amount of $Q$ per loop in the trajectory is the product of the area $S$ of the loop and $E_{Cm}$. In the most frequent case, $S$ = 0 and thus $Q$ = 0. (d) Schematic illustration of the decomposition of the trajectory of [$N_1(t)$, $N_2(t)$] into closed loops. The trajectory is decomposed into five loops: one with $Q = E_{Cm}$, one with $Q = -3E_{Cm}$, and three with $Q$ = 0. (e) The probability distribution of $Q$, $P(Q)$, derived by the trajectory obtained from the experiment (orange-filled bins) and a Monte Carlo simulation (blue line). As the capacitively coupled dots are at equilibrium, $P(Q)$ is symmetric and the net value of $Q$ is zero. The Monte Carlo simulations were performed with $dt$ = 0.1 s, $t_{meas}$ = 30,000 s, $\Gamma_{P1}$ = 0.39 s$^{-1}$, $\Gamma_{P2}$ = 0.15 s$^{-1}$, $E_{C1}$ = 7 meV, $E_{C2}$ = 9 meV, $kT$ = 26 meV, and $E_{Cm}$ = 2.0 meV.